# Multifunctional Lateral Transition-Metal Disulfides Heterojunctions

Yipeng An,[*] Yusheng Hou, Kun Wang, Shijing Gong, Chunlan Ma, Chuanxi Zhao, Tianxing Wang, Zhaoyong Jiao, Heyan Wang, and Ruqian Wu[*]

**Abstract**

The intrinsic spin-dependent transport properties of two types of lateral $VS_2|MoS_2$ heterojunctions are systematically investigated using first-principles calculations, and their various nanodevices with novel properties are designed. The lateral $VS_2|MoS_2$ heterojunction diodes show a perfect rectifying effect and are promising for the applications of Schottky diodes. A large spin-polarization ratio is observed for the A-type device and pure spin-mediated current is then realized. The gate voltage significantly tunes the current and rectification ratio of their field-effect transistors (FETs). In addition, they all have sensitive photoresponse to blue light, and could be used as photodetector and photovoltaic device. Moreover, they generate the effective thermally-driven current when a temperature gratitude appears between the two terminals, suggesting them as potential thermoelectric materials. Hence, the lateral $VS_2|MoS_2$ heterojunctions show a multifunctional nature and have various potential applications in spintronics, optoelectronics, and spin caloritronics.

Prof. Y. P. An, Prof. T. X. Wang, Prof. Z. Y. Jiao, H. Y. Wang
School of Physics & Henan Key Laboratory of Boron Chemistry and Advanced Energy Materials, Henan Normal University, Xinxiang 453007, China
E-mail: ypan@htu.edu.cn
Dr. Y. S. Hou, Prof. R. Q. Wu
Department of Physics and Astronomy, University of California, Irvine, California 92697, USA
E-mail: wur@uci.edu
Dr. K. Wang
Department of Mechanical Engineering, University of Michigan, Ann Arbor, MI 48109 USA
Prof. S. J. Gong
Key Laboratory of Polar Materials and Devices (MOE) & Department of Optoelectronics, East China Normal University, Shanghai 200062, China
Prof. C. L. Ma
School of Mathematics and Physics, Suzhou University of Science and Technology, Suzhou, Jiangsu 215009, China
Prof. C. X. Zhao
Siyuan Laboratory, Guangdong Provincial Engineering Technology Research Center of Vacuum Coating Technologies and New Energy Materials, Department of Physics, Jinan University, Guangzhou 510632, China

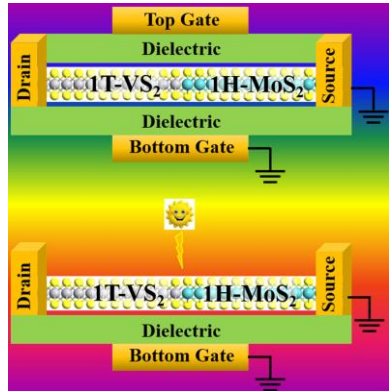

# 1. Introduction

After the discovery of the well-known graphene,[1] a large number of two-dimensional (2D) monolayers (MLs) have been fabricated and studied, including silicene,[2,3] h-BN,[4] transition metal sulfides,[5-10] phosphorene,[11-15] MXene,[16,17] borophene,[18-23] stanene,[24] antimonene,[25] g-$C_3N_4$,[26,27] ferromagnets,[28,29] chromium trihalides $CrX_3$,[30-33] and other graphene-like MLs.[34-37] Each type of these 2D materials possesses unique properties in mechanics, thermodynamics, optics, electrics, and magnetism. Novel properties can arise when different 2D materials are combined vertically to form a van der Waals (vdW) heterojunction or via in-plane stitching to form a lateral heterojunction.[38-40] A few lateral or vdW heterojunctions have been investigated theoretically or prepared experimentally, such as graphene|h-BN,[41] graphene|$MoS_2$,[42] borophene|organic lateral heterojunctions,[43] and InSe vdW heterojunction,[44] etc.

In particular, the transition-metal disulfides (TMD) are among the most explored 2D materials. Geometrically, the TMD mostly exist in 1T ($D_{3D}$), 2H ($D_{3h}$), and 3R ($D_{3v}$), based on their distinct symmetries.[45] Electronically, they may be semiconductors, insulators, and metals. In recent years, some lateral and vdW heterojunctions of TMD with same or different phases have been realized experimentally or investigated theoretically, including MoS$_2$|WSe$_2$,[46-49] MoSe$_2$|WSe$_2$,[50,51] NbS$_2$|WS$_2$,[52] and VSe$_2$|MX$_2$.[53] These heterojunctions hold promise for potential applications as transistors (including diodes and triodes), photoelectric devices such as photodetectors, etc. Very recently, the high-quality lateral metal-semiconductor heterostructures VS$_2$|MoS$_2$ were prepared via a two-step chemical vapor deposition (CVD) method,[54] and were proven to have better field-effect mobility compared to the traditional on-top Ni contacts.

As the ground state of 1T phase VS$_2$ ML is ferromagnetic,[55] unusual spin-dependent properties could arise from the lateral VS$_2$|MoS$_2$ heterojunctions, making them interesting candidates for spintronic applications. A detailed study for this issue is still missing. In this paper, we systematically investigate the spin-dependent transport and photoelectric properties of the lateral VS$_2$|MoS$_2$ heterojunctions by the first-principles calculations. We first construct the lateral VS$_2$|MoS$_2$ heterojunction diodes and study their intrinsic spin-dependent electronic transport properties. Subsequently, their field-effect transistors are obtained by adding the double (bottom and top) gate electrodes on both sides of the heterojunction, and the field-effect properties are then unveiled. Next, we further investigate their photoelectric properties as they are under illumination. Last, we explore their thermally-driven electron transport behaviors due to a temperature difference between their two terminals.

## 2. Results and Discussion

Based on our first-principles calculations, the ground state of 1T-VS$_2$ ML is a ferromagnetic metal (0.55 $\mu_B$ per unit cell) with 0.02 eV lower energy than its spin-unpolarized configuration, while the 1H-MoS$_2$ ML is a semiconductor with a gap of 1.7 eV (see Supporting Information Figure S1), consistent with the previous reports.[55-57]

According to the atomic lattice structures of 1T- and 1H-phase transition-metal disulfides, there exist two types of in-plane stitching patterns for 1T-VS$_2$ and 1H-MoS$_2$ to form a lateral heterojunction. The first type is along the zigzag direction (labeled as Z-type), and the second type is along the armchair direction (labeled as A-type), as shown in Figure S2 of the Supporting Information. These two types of lateral VS$_2$|MoS$_2$ heterojunctions may yield distinct spin-dependent transport behavior since the band structures of either 1T-VS$_2$ or 1H-MoS$_2$ are significantly different between along the $\Gamma-X$ path (i.e., parallel to the transport direction of Z-type) and along the $\Gamma-Y$ path (i.e., parallel to the transport direction of A-type), as shown in the Supporting Information Figures S1e and S1f.

## 2.1. Intrinsic rectifying effect of VS$_2$|MoS$_2$ Schottky diodes

Figure 1 shows the schematics of the two types of lateral VS$_2$|MoS$_2$ heterojunction diodes. Each diode is composed of the periodic drain (D) and source (S) electrodes and the central scattering region VS$_2$|MoS$_2$ heterojunction. They have a periodicity perpendicular to the transport direction between the D and S electrodes. The D/S electrodes are described by a large supercell of VS$_2$ or MoS$_2$, and their length are semi-infinite along the transport direction. The third direction is a large slab separated by a vacuum more than 15 Å. The Dirichlet boundary conditions is adopted along the transport direction, while periodic boundary conditions for the other two orthogonal directions. As a D-S bias $V_b$ is applied across the D electrode, its Fermi energy ($E_F$) is shifted upward (for forward $V_b$) or downward (for reverse $V_b$) accordingly. A forward $V_b$ generates a current from the D electrode to the S electrode, and vice versa. In this work, the spin-resolved currents through the VS$_2$|MoS$_2$ heterojunctions are calculated using the Landauer-Büttiker formalism[58]

$$I_\sigma(V_b) = \frac{e}{h} \int_{\mu_D}^{\mu_S} T^\sigma(E, V_b)[f_D(E-\mu_D) - f_S(E-\mu_S)] \mathrm{d}E, \qquad (1)$$

where the $\sigma = \uparrow$ (spin up) and $\downarrow$ (spin down), and the total current $I$ is the sum of $I_\sigma$. $e$ and $h$ are the electron charge and the Planck's constant, respectively. $T^\sigma(E, V_b)$ indicates the spin-resolved transmission coefficient of the heterojunctions. $f_{D(S)} = \{1 + \exp[(E - \mu_{D(S)})/k_B T_{D(S)}]\}^{-1}$ is the Fermi-Dirac distribution function of the D(S) electrode with temperature $T_{D(S)}$ and chemical potential $\mu_{D(S)}$.

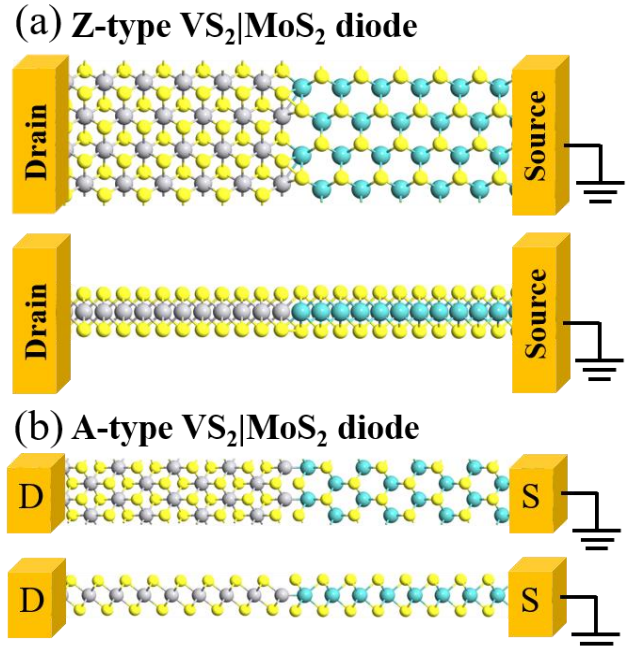

**Figure 1.** Schematics of top (upper) and side (lower) views of a) Z-type and b) A-type lateral $VS_2|MoS_2$ heterojunction diodes.

Figure 2a,b show the spin up and spin down (dn) transmission spectra of Z-type $VS_2|MoS_2$ diode. Only the electrons with high enough negative energies have nonnegligible transmission coefficients for the both spin states. More than that, within the bias window, significant electron transmissions only appear under high enough forward bias voltages for both spin states. Besides, the spin down states show a slightly stronger transmission ability than the spin up sates. These effects result in a perfect rectifying effect of Z-type $VS_2|MoS_2$ heterojunction (see Figure 2e), which is promising for the applications of Schottky diodes. The current through the heterojunction is turned on as the bias is larger than the forward

threshold voltage (0.4 V). Its total current density and differential conductance density are as high as 670 mA/mm and 3.6 S/mm (see Figure 2g) at 0.8 V, respectively. Note that a disorder or mismatch (such as vacancy defect) may appear at the interface when preparing a lateral heterojunction. It usually reduces the electrical conductivity of 2D materials because the vacancy defect decreases the electron transmission channels and increases the electron scattering.[21,59] The vacancy defect such as from V atom decreases the conductivity but remains the rectification effect of Z-type VS$_2$|MoS$_2$ with reduced rectification ratio (see Supporting Information Figure S3).

Another important parameter for a rectifier is the rectification ratio defined as RR = $|I(V_b)/I(-V_b)|$. The RR of Z-type VS$_2$|MoS$_2$ heterojunction is up to the magnitude of $10^7$, much larger than that of the graphene|h-BN heterojunction[60] and other nano rectifiers.[61,62] Moreover, according to the thermionic emission theory,[63] the $I-V$ characteristics of the Z-type VS$_2$|MoS$_2$ Schottky diode can be described as

$$I = I_0 e^{qV_b/nk_BT}(1 - e^{-qV_b/k_BT}) \qquad (2)$$

where $q$ is the elementary charge, $T$ indicates the temperature, $I_0$ is the saturation current, and $n$ refers to the so-called ideality factor. The ideality factor is a key parameter to evaluate how much the heterojunction resembles an ideal Schottky diode, where a value of $|n|=1$ represents the idea case. Using a log-plot of $I/(1 - e^{-qV_b/k_BT})$ against $V_b$, one can extract the value of the ideality factor from the slope. The ideality factor of Z-type VS$_2$|MoS$_2$ Schottky diode is 0.94 at room temperature (see Figure 2i), close to a saturation value (0.91) as the temperature is beyond 360 K (see Figure 2j). The temperature for the idea case of Z-type VS$_2$|MoS$_2$ Schottky diode ($n = 1$) is 275 K.

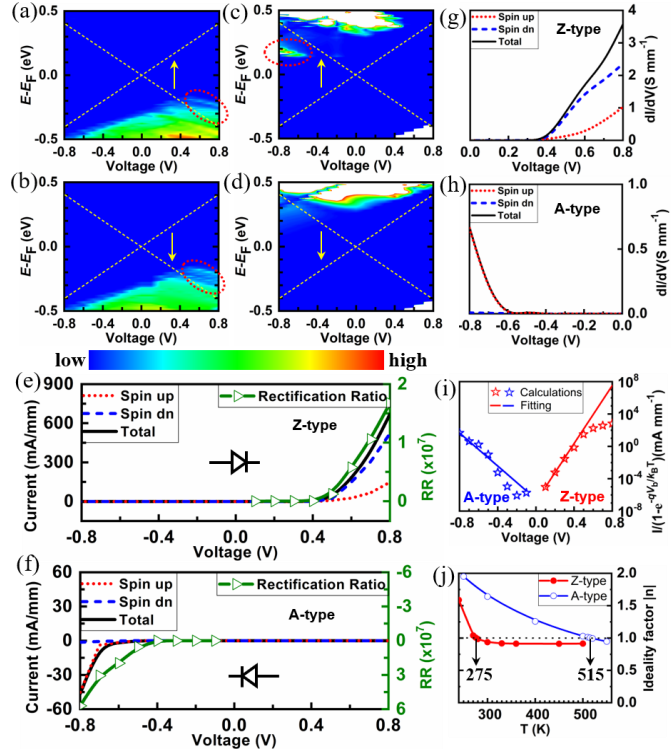

**Figure 2.** Intrinsic a) spin up and b) spin down transmission spectra of Z-type VS$_2$|MoS$_2$ diode. c) spin up and d) spin down transmission spectra of A-type VS$_2$|MoS$_2$ diode. The yellow dotted lines refer to the bias window. The ↑ and ↓ indicate the spin up and spin down, respectively. The intrinsic spin-resolved $I-V$, rectification ratio, and dI/dV curves of VS$_2$|MoS$_2$ heterojunction diodes, e) and g) for Z-type, f) and h) for A-type. (i) $I/(1 - e^{-qV_b/k_BT})$ against $V_b$ curve and j) Temperature-dependent ideality factor |n| curve of VS$_2$|MoS$_2$ heterojunction diodes. The dotted line in j (|n| = 1) indicates the idea case.

The A-type diode also shows a Schottky diode behavior, and has large transmission ability for the spin-polarized electrons under the region of high positive energies (see Figure 2c,d). Within the bias window, only the spin up electrons have significant transmission and they appear at the reverse bias region, while there is no effective transmission for the spin down electrons. In addition, the A-type diode exhibits a perfect spin filtering effect and can be utilized as a spin filter (see Figure 2f). For instance, the spin-polarization ratio, defined as SP = (I$_↑$−I$_↓$)/(I$_↑$+I$_↓$), is as high as 94% at −0.8 V bias, close to a VS$_2$|MoS$_2$|VS$_2$ magnetic tunnel

heterojunction [64] and larger than a 2D phosphorene monolayer.[65] The open-circuit voltage of A-type Schottky diode is −0.6 V. Its total current density and differential conductance density are 47 mA/mm and 0.7 S/mm (see Figure 2f,h) at −0.8 V, smaller than Z-type VS$_2$|MoS$_2$ diode. Moreover, the ideality factor of A-type VS$_2$|MoS$_2$ Schottky diode is −1.64 at room temperature (see Figure 2i). Its absolute value decreases gradually as the temperature increases (see Figure 2j), and reaches to the idea case at 515 K. Therefore, the A-type VS$_2$|MoS$_2$ diode could also become a Schottky diode device with considerable rectification ratio as the Z-type does.

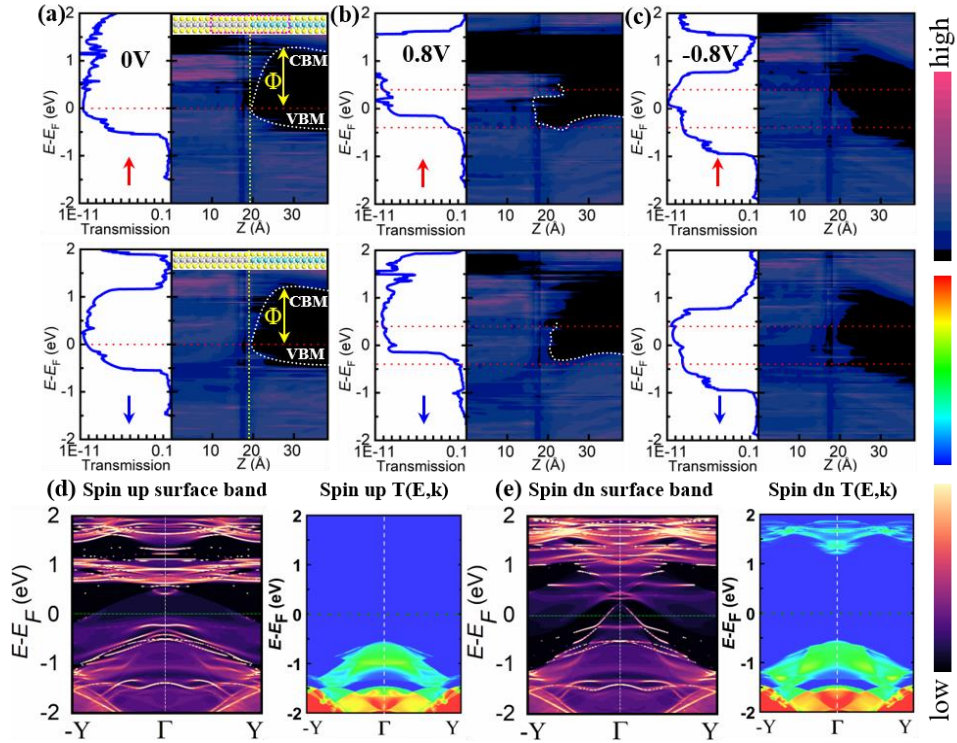

**Figure 3.** Projected local density of states of Z-type VS$_2$|MoS$_2$ heterojunction at a) 0, b) 0.8, and c) −0.8 V. The upper (lower) refers to the spin up (down) PLDOS. The Fermi level is set to zero. Φ is the energy barrier. CBM and VBM indicate the conduction band minimum and valence band maximum values, respectively. The red dotted lines refer to the bias window. Surface band (spectral function) of the region near the interface (see the pink rectangle of Figure 3a) and k-dependent transmission coefficients for the spin up (d) and spin down (e) states.

To understand the outstanding rectifying effect of the two types of VS$_2$|MoS$_2$ heterojunctions, we further calculate and analyze their projected local density of states (PLDOS) under the biases of 0, 0.8, and −0.8 V, respectively. For the Z-type of VS$_2$|MoS$_2$ heterojunction, at 0 V bias (see Figure 3a), there is little electron transmission near the $E_F$ for either the spin up states or the spin down states. This is because that the 1H-MoS$_2$ part of the heterojunction has a high energy barrier Φ and large energy gap, according to their PLDOS. This electronic structure coupling and band alignment can be well understood by their band structures (see Supporting Information Figures S1e and S1f). Figure 3d,e show the surface band (spectral function) of the region near the interface and k-dependent transmission coefficients. The both spin states have a strong distribution near the Γ point for either the surface band or T(E, k). Compared to the spin up state, the surface band of the spin down state is more delocalized in the positive energy. This contributes to the additional spin-down electron transmissions within the corresponding positive energy region.

As the forward bias increases towards 0.8 V (see Figure 3b), the energies of 1T-VS$_2$ part shifts down, and the spin down states can better facilitate electron transmission within the expanded bias window. However, under the reverse bias such as −0.8 V (see Figure 3c), the energies of 1T-VS$_2$ part shifts up, leading to a large energy gap of 1H-MoS$_2$ part covering the bias window. These features result in the rectifying effect of Z-type VS$_2$|MoS$_2$ heterojunction with perfect rectification ratios and a better spin down channel. For the case of A-type VS$_2$|MoS$_2$, it has the same rectifying mechanism (see Supporting Information Figure S4), but its rectifying behavior appears under the reverse bias region (see Figure 2f).

## 2.2. Field effect transistors based on the VS$_2$|MoS$_2$ heterojunctions

Next, we construct the field-effect transistor (FET) structures for the two types of VS$_2$|MoS$_2$ heterojunctions (see Figure 4a,b), and investigate their spin-dependent field-effect properties. Both the bottom and top gates are positioned near the central interface region of

each heterojunction. We recalculate their $I-V$ curves from −0.8 to 0.8 V at different gate voltages.

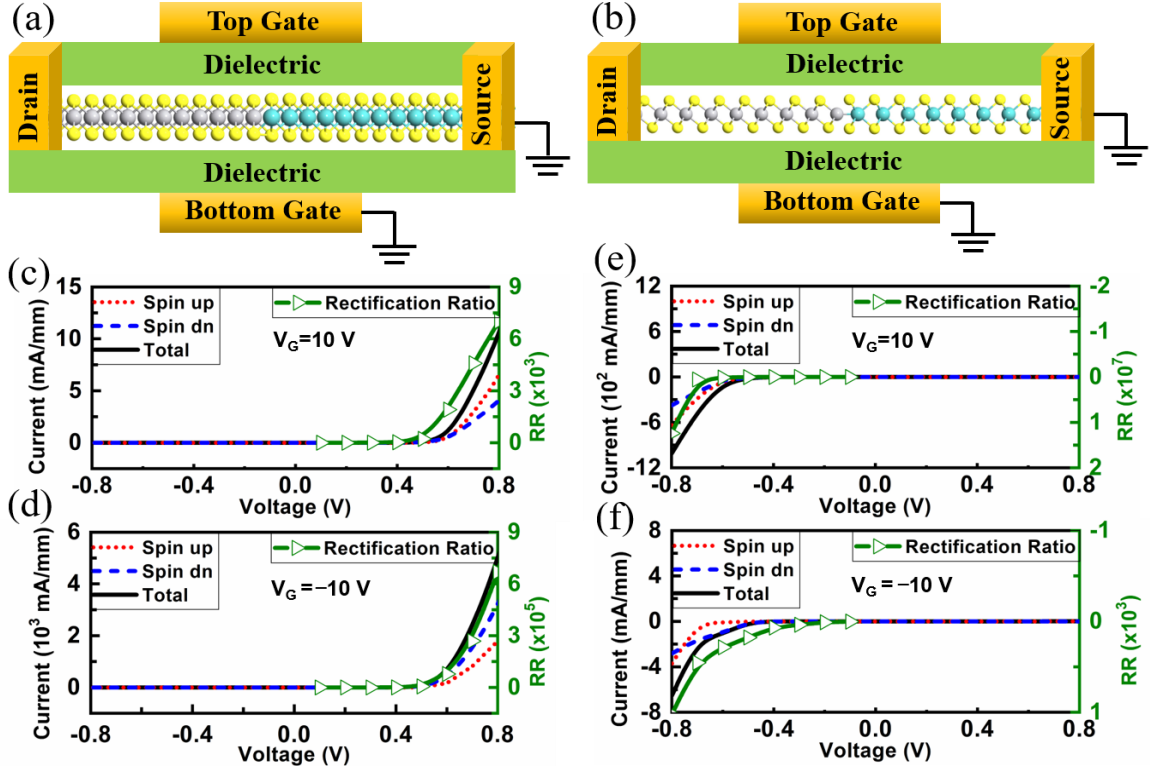

**Figure 4.** Schematics of two types of VS$_2$|MoS$_2$ field-effect transistors and their $I-V$ curves at different gate voltages. a), c), and d) for Z-type. b), e), and f) for A-type.

Figure 4c,d show the $I-V$ curves of Z-type VS$_2$|MoS$_2$ FET at the gate voltages of 10 and −10 V, respectively. The positive gate voltage drastically decreases the spin up and spin down current, as well as the total current. Simultaneously, the rectification ratios reduce by four orders of magnitude to $10^3$. This is because that the positive gate voltage shifts down the energy of the heterojunction, which moves the $E_F$ to the energy gap central of the 1H-MoS$_2$ part (see Figure 3). While for the case of −10 V gate voltage, the currents increase by sevenfold, and the rectification ratio reduces by five order of magnitude. This moment the negative gate voltage shifts the energy up, causing more PLDOS to enter into the bias window (see Figure 3a). This effect leads to the increase in currents. In addition, the threshold voltage for both the spin up and the spin down states increases to 0.5 V from 0.4 V.

For the A-type VS$_2$|MoS$_2$ FET, the positive gate voltage (10 V) turns on the spin down channel, leading to a current increase by twenty-two times (see Figure 4e) while keeping the rectification ratio unchanged. This is because more conductive band states enter into the bias window under a positive gate voltage (see Supporting Information Figures S4). While under a negative gate voltage (−10 V), some conductive band states shift out the bias window. This causes a slight current decrease (see Figure 4f) and leads to the rectification ratio decrease by three order of magnitude. Note that both gate voltage polarities lead to a significant reduction in spin-polarization ratio.

## 2.3. Intrinsic photoelectric properties of VS$_2$|MoS$_2$ heterojunctions and their phototransistors

Various recent studies have shown 2D materials hold great potential for applications as photoelectric nanodevices.[66-69] For instance, phosphorene and palladium diselenide can be utilized as the photodetectors.[68] Graphene-based heterojunctions and black phosphorus have been shown to play a significant role in infrared and mid-infrared photodetectors.[70-72] To further investigate the photoelectric properties of VS$_2$|MoS$_2$ heterojunctions, we construct their transistor structures (see Figure 5a,b) and study their intrinsic photocurrents under illumination, as well as the regulation effect due to the gate electrode. In this work, the linearly polarized light along the transport direction is employed and the incident photon energy is from 0 to 5 eV. The photocurrent through the transistors can be obtained as a first-order perturbation to the electronic system which stems from the interaction with a weak electromagnetic field. The electron-photon interaction of the whole system is given by the Hamiltonian

$$H^{'} = \frac{e}{m_0} \mathbf{A} \cdot \mathbf{P}, \tag{3}$$

where **A** and **P** are the vector potential and momentum operator, respectively. Due to absorption of $N$ photons with frequency $\omega$, the photoexcited currents into electrode $\alpha$ = D/S is obtained by

$$I_\alpha = \frac{e}{h}\int_{-\infty}^{\infty}\sum_{\beta=D,S}[1-f_\alpha(E)]f_\beta(E-\hbar\omega)T_{\alpha,\beta}^{-}(E) - f_\alpha(E)[1-f_\beta(E+\hbar\omega)]T_{\alpha,\beta}^{+}(E)\mathrm{d}E. \quad (4)$$

The total photocurrent is then given by $I_{ph} = I_D - I_S$. More details can be found in the previous reports.[73-75]

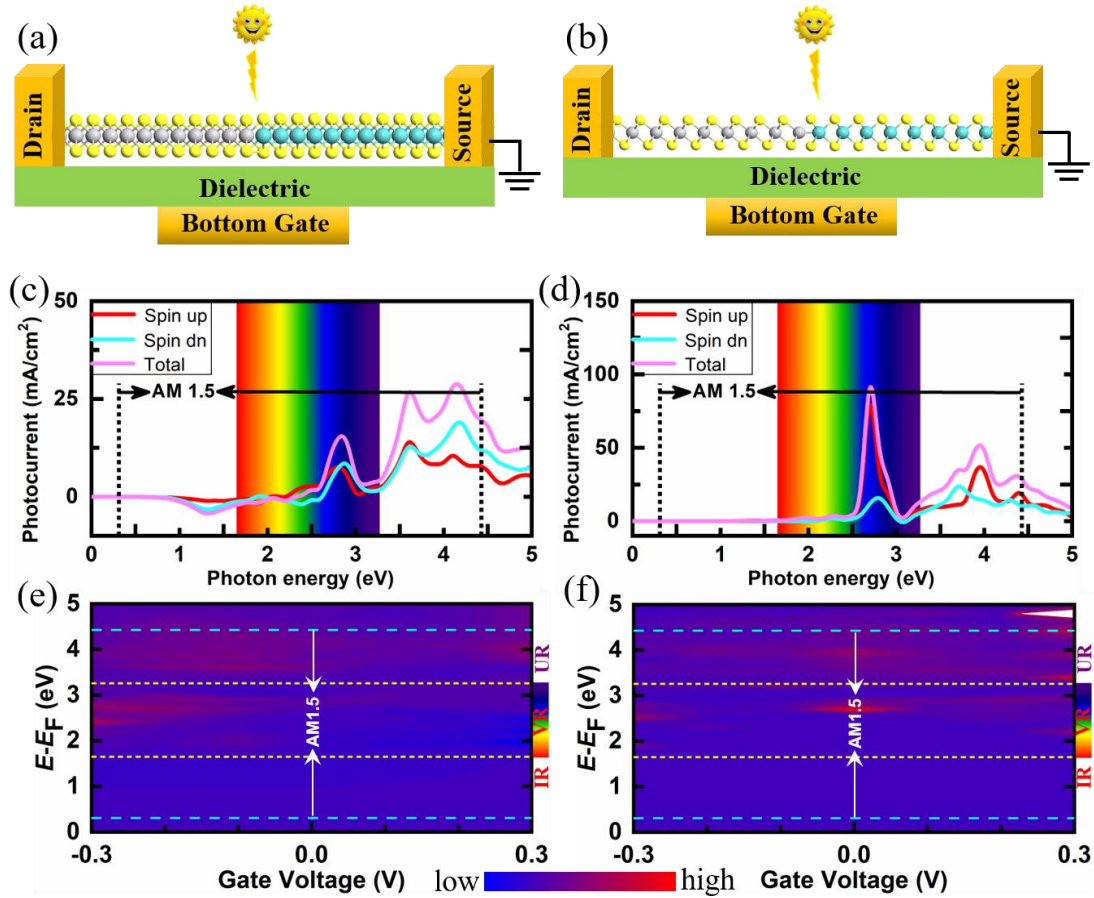

**Figure 5.** Schematics, intrinsic photocurrent, and gate-dependent photocurrent spectra of VS$_2$|MoS$_2$ photoelectric devices. a), c), and e) for Z-type. b), d), and f) for A-type. IR, VR, and UR in e) and f) refer to the infrared-light region, visible-light region, and ultraviolet-light region, respectively.

Figure 5c shows the intrinsic spin-dependent photocurrent curves of Z-type VS$_2$|MoS$_2$ heterojunction under illumination. Under the infrared-light region (IR), the photoexcited

current is very small because the low photon energy cannot overcome the large energy barrier of the 1H-MoS$_2$. Within the visible-light region (VR), this heterojunction is highly sensitive to blue light as a photocurrent peak can be seen in the area for the spin up, spin down, and the total values. Hence, this Z-type VS$_2$|MoS$_2$ heterojunction can be used as a photodetector for blue light detection. Moreover, the photocurrents become large at the ultraviolet region (UR) because the photons with high energy open up more transition channels. Within the AM1.5 standard,[76] the total induced photocurrent density per second is 29.6 mA/cm$^2$, which is three orders of magnitude higher than that of hydrogenated borophene.[20] This demonstrates that the Z-type VS$_2$|MoS$_2$ heterojunction could also lead to applications as photovoltaic devices. For the case of A-type VS$_2$|MoS$_2$ heterojunction, it displays similar photoelectric properties, but excites a slightly larger total photocurrent (34.2 mA/cm$^2$) within the AM1.5 range (see Figure 5d).

We further investigate the three-terminal phototransistor properties of VS$_2$|MoS$_2$ heterojunctions by adding a gate electrode at the bottom to tune their photoelectric properties. Figure 5e shows the photocurrent spectra of Z-type VS$_2$|MoS$_2$ transistor under the gate voltages from −0.3 to 0.3 V. Under the AM1.5 standard, the negative gate voltages can enhance the photoexcited current except the IR. Particularly, near the blue light region of VR, the photocurrent is significantly enhanced under the negative gate voltages. For the A-type of VS$_2$|MoS$_2$ phototransistor, positive gate voltages lead to greater increase in photocurrents within the VR and UR (see Figure 5f).

## 2.4. Thermospin diode based on the lateral VS$_2$|MoS$_2$ heterojunctions

Traditional spintronics mainly focus on the coupled electron charge and spin transport in condensed-matter structures and devices. The rapidly growing research field of spin caloritronics is devoted to utilizing the excess heat from the nanoscale materials and devices to excite the flow of spin-mediated current.[77,78] The thermally-driven spin-mediated current is the central to the spin Seebeck effect (SSE), which has been regarded as a promising approach to generate pure spin-mediated current via creating thermal gradient across magnetic heterojunctions. It has extensive utilizations in thermoelectric materials and devices.[79,80] So, we further explore the thermally-driven spin transport properties of lateral $VS_2|MoS_2$ heterojunctions.

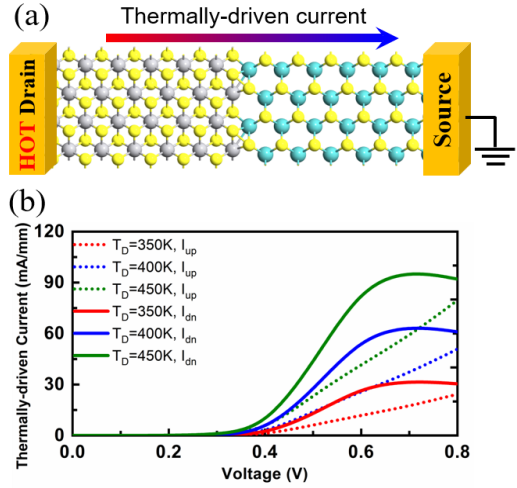

**Figure 6.** a) Schematic of Z-type lateral $VS_2|MoS_2$ heterojunction thermospin diodes. b) Bias-dependent thermally-driven spin-mediated current under various drain temperatures ($T_D$).

Figure 6a shows the schematic of Z-type lateral $VS_2|MoS_2$ heterojunction thermospin diode. Due to a limited temperature gradient $\Delta T$ (= $T_D - T_S$) arising across the drain and the source electrodes, a thermally-driven current $I_{th}$ may be then generated in a linear-response fashion using the Equation 1. The difference in carrier concentrations between the drain and the source depends on the Fermi distribution ($f_D - f_S$), which is closely related to the electron temperature of the two electrodes. Due to a large energy gap of the 1H-$MoS_2$, the temperature change of the source has no significant impact on the Fermi distribution. Hence, we study the thermally-driven current effect of the Z-type lateral $VS_2|MoS_2$ heterojunction thermospin diode, by fixing the source to the room temperature and increasing the drain temperature.

Figure 6b shows the bias-dependent thermally-driven spin-mediated current at various drain temperatures. The thermally-driven currents $I_{th}$ of both the spin up and spin down states are turned on since the bias is beyond the threshold voltages 0.4 V, which is consistent with its rectifying diode. In addition, the $I_{th}$ increases gradually with the increase of drain temperature, due to the enhanced thermal broadening of the Fermi distribution. The difference is only that, the $I_{th}$ for the spin up (down) states is nonlinear (linear). This may originate from the different spin-polarized band structures of 1T-VS$_2$ along the $\Gamma-X$ path (see Supporting Information Figure S1e). Therefore, the Z-type lateral VS$_2$|MoS$_2$ heterojunction thermospin diode can generate large thermally-driven current and be potentially utilized as thermoelectric materials and devices. Note that the magnitude of the thermally-driven spin-mediated current of the A-type lateral VS$_2$|MoS$_2$ heterojunction is small due to the larger energy gap of 1H-MoS$_2$ along the $\Gamma-Y$ path.

## 3. Conclusion

In summary, the intrinsic spin transport properties of lateral VS$_2$|MoS$_2$ heterojunctions are systematically studied by means of first-principles calculations. We demonstrate various device structures and their novel transport properties. The Z- and A-type lateral VS$_2$|MoS$_2$ heterojunction diodes show a perfect rectifying effect at the forward and reverse bias voltages, rendering them promising candidates as Schottky diodes. Both of them present a spin-polarization behavior, with the A-type heterojunction showing a large spin-polarization ratio as high as 94% at −0.8 V. For their FET devices, the positive (negative) gate voltages decrease (increase) in current for Z-type (A-type). Their rectification ratios are generally decreased under a nonzero gate voltage. Additionally, under the illumination, both the two types of lateral VS$_2$|MoS$_2$ heterojunctions have sensitive photoresponse to blue light within

the visible-light, and thus can be utilized as a photodetector. They can generate exploitable photocurrents under the AM1.5 standard and could be a good candidate for photovoltaic applications. What's more, a thermally-driven current can appear due to a temperature difference between the two terminals of the heterojunction device. This feature makes them possible materials for thermoelectric energy conversion applications. Therefore, the lateral metal-semiconductor heterojunctions of transition-metal disulfides, $VS_2|MoS_2$, hold promise for developing various applications in spintronics, optoelectronics, and spin calorimetry.

## 4. Experimental Section

All the first-principles self-consistent calculations in this work are performed by using the density-functional theory and non-equilibrium Green's function method (NEGF-DFT), as implemented in the Atomistix Toolkit code.[81-83] The exchange and correlation effects of electrons are described with the spin-polarized GGA-PBE functional.[84,85] The core electrons of all atoms are represented by the norm-conserving Vanderbilt (ONCV) pseudopotentials.[86] Linear combinations of atomic orbitals (LCAO) basis sets are used to expand the wave-functions of valence electrons. A real-space grid density which is equivalent to a plane-wave kinetic energy cutoff of 100 Ha is used. The Monkhorst-Pack $k$-point grids $1 \times 5 \times 300$ and $1 \times 9 \times 200$ are adopted to sample the Brillouin zone for the electrodes of Z-type and A-type $VS_2|MoS_2$ heterojunctions, respectively. The total energy tolerance and residual force on each atom are less than $10^{-6}$ eV and 0.01 eV/Å in the geometry optimization, respectively.

The self-consistent electronic structures of all these device systems can be determined by solving the Kohn-Sham and Poisson equations within two different boundary conditions (periodic and Dirichlet).[87,88] It is periodic for the electrode regions along the transport direction and the whole device structure along another perpendicular directions. A Dirichlet boundary condition is adopted at the boundary between the electrode and the central scattering region. The sample width (central scattering region) is 3.8 and 4.4 nm for the Z-type and A-type VS$_2$|MoS$_2$ transistors, respectively. The metallic gate is 2.0 nm wide and located near the interface region. The dielectric constant of the substrate is set to $4.0\,\varepsilon_0$, close to that of SiO$_2$.

**Supporting Information**

Supporting Information is available from the Wiley Online Library or from the author.

# Acknowledgements

The work at HNU was supported by the NSFC (Grants No. 11774079, No. 61774059 and No. U1704136), the Scientific and Technological Innovation Program of Henan Province's Universities (Grant No. 20HASTIT026), the Young Backbone Teacher Program of Henan Province's Higher Education (Grant No. 2017GGJS043), the Science Foundation for the Excellent Youth Scholars of Henan Province (2020 year), the Henan Overseas Expertise Introduction Center for Discipline Innovation (Grant No. CXJD2019005), the Science Foundation for the Excellent Youth Scholars of HNU (Grant No. 2016YQ05), and the HPCC of HNU. The work at UCI was supported by the US DOE-BES (Grant No. DE-FG02-05ER46237). We thank W. Ju and D. Kang at HNUST for helpful discussion.

**Conflict of Interest**

The authors declare no conflict of interest.

**Table of contents entry:**

The intrinsic spin-resolved transport properties of two types of lateral transition-metal disulfides $VS_2|MoS_2$ heterojunctions are unveiled, and their various multifunctional nanodevices are designed. The lateral $VS_2|MoS_2$ heterojunctions are promising for the applications of rectifying diodes with a perfect rectification ratio, spin filters with large spin-polarization ratio, field-effect transistors, photodetectors, photovoltaic devices, and thermoelectric materials.

**Keywords:**

Lateral heterojunctions, transition-metal disulfides, nanodevices, Schottky diodes, field-effect transistors.

**Authors:**

Yipeng An,[*] Yusheng Hou, Kun Wang, Shijing Gong, Chunlan Ma, Chuanxi Zhao, Tianxing Wang, Zhaoyong Jiao, Heyan Wang, and Ruqian Wu[*]

**Title:**

Multifunctional Lateral Transition-Metal Disulfides Heterojunctions

**ToC figure:**

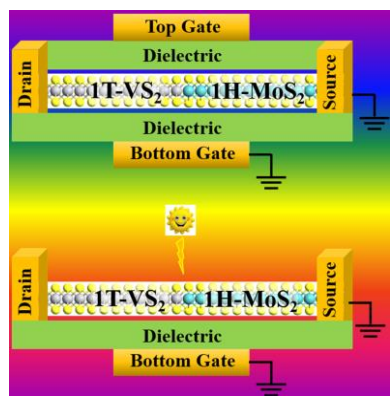